\documentclass[footinbib,twocolumn,showpacs,amsmath,amstex,amssymb,mathfonts,superscriptaddress,prb]{revtex4}
\usepackage{graphicx}
\usepackage{dcolumn}
\usepackage{float}
\usepackage{amsmath}
\usepackage{bm}
\usepackage{array}
\usepackage{multirow}
\usepackage{hyperref}
\usepackage{amsmath}
\usepackage{amssymb}
\usepackage{amsthm}
\usepackage{amsfonts}
\usepackage{epstopdf}

\begin{document}

\title{The behavior of the fractional quantum Hall states in the LLL and 1LL with in-plane magnetic field and Landau level mixing: a numerical investigation}
\author{Lin-Peng Yang} 
\author{Qi Li} 
\author{Zi-Xiang Hu}
\email{zxhu@cqu.edu.cn}
\affiliation{Department of Physics, Chongqing University, Chongqing, P.R. China, 401331.}
\date{\today}

\begin{abstract}
By exactly solving the effective two-body interaction for two-dimensional electron system with layer thickness and an in-plane magnetic field, we recently found that the effective interaction can be described by the generalized pseudopotentials (PPs) without the rotational symmetry. With this pseudopotential description, we numerically investigate the behavior of the fractional quantum Hall (FQH) states both in the lowest Landau level (LLL) and first Landau level (1LL).  The enhancements of the $7/3$ FQH state on the 1LL for a small tilted magnetic field are observed when layer thickness is larger than some critical values. While the gap of the $1/3$ state in the LLL monotonically reduced with increasing the in-plane field.  From the static structure factor calculation, we find that the systems are strongly anisotropic and finally enter into a stripe phase with a large tilting. With considering the Landau level mixing correction on the two-body interaction, we find the strong LL mixing cancels the enhancements of the FQH states in the 1LL.
\end{abstract}

\maketitle 
\section{Introduction}
A two-dimensional electron gas system in a strong perpendicular magnetic field displays a host of collective ground states. The underlying reason is the formation of two-dimensional Landau levels (LLs) in which the kinetic energy is completely quenched. In a macroscopically degenerate Hilbert space of a given LL, the Coulomb potential between electrons is dominated which makes the system strongly interacting. The different fractional quantum Hall (FQH) states are realized~\cite{tsui, prange} for different specific interactions. For example,  in the description of the Haldane's PPs~\cite{haldane1, haldane2} for interaction with rotational symmetry, the celebrated Laughlin~\cite{laughlin} states at filling fraction $\nu = 1/3$ and $\nu = 1/5$ are dominated by $V_1$ and $V_1, V_3$ potentials respectively where $V_m$ is the component in the effective interaction with two electrons having relative angular momentum $m$.  Naturally, the FQH state will be destroyed while these dominant PPs are weaken by the realistic interaction in the systems, i.e. the increase of the $V_m(m>1)$ can decrease the  $V_1$.  Therefore, the analysis of the effective interactions should be very helpful to investigate the properties of the FQH liquids.

The interaction between electrons can be tuned by various choices of ``experimental knobs", such as the layer thickness of the sample, the effects of the LL mixing from unoccupied LLs,  the in-plane magnetic field and other external methods such as lattice strain or electric field.   With considering these effects on the ideal electron-electron interaction, some of the inherent symmetry will be broken. For example, the LL mixing breaks the particle-hole symmetry which makes the $5/2$ state on the 1LL to be mysterious for more than two decades~\cite{MR, Wilczek91}. The in-plane magnetic field~\cite{lilly99,du99,dean08,chi12} for electrons or the in-plane component for dipolar fermions~\cite{Cooper08, qiu11} introduce anisotropy and break the rotational symmetry of the system.  In the absence of the rotational invariance~\cite{haldane3}, we recently~\cite{yang2} generalized the pseudopotential description without conserving the angular momentum in which the anisotropy of the system is depicted by non-zero non-diagonal PPs $V_{m,n}(m \neq n)$.  Some of the anisotropic interactions can be simply modeled by few PPs. For instance we found the anisotropic interaction of the dipolar fermions in FQH regime can be modeled by $V_1 + \lambda V_{1,2}$ in the LLL and $V_1 + V_3 + \lambda_1 V_{1,2} + \lambda_2 V_{3,2}$ in the 1LL~\cite{hu18}.  The effect of anisotropy introduced by an in-plane magnetic field has been quite consistent in the LLL. In the $\nu = 1/3$ Laughlin state, the incompressibility generally decreases while increasing the anisotropy of the system~\cite{zpapic13, yang1}. However,  people found that the experimental results on the 1LL are controversial. Different experimental results are observed in different samples~\cite{dean08, chi12}.  Some of them reveal that increasing the in-plane B field stabilize the FQH states and some of them destabilize the FQH states. Therefore, for the imcompressible states, one needs a more accurate theoretical prediction on how the stability of the FQH states in 1LL (such as FQH states at $\nu =7/3, 8/3$ and $\nu = 5/2, 7/2$ et.al.) in the presence of a tilted magnetic field. It was suggested that the width of the quantum well plays an important role in explaining these different results.~\cite{dean08}

 In this paper, based on the pseudoptential description of the electron-electron interaction in a titled magnetic field with a finite layer thickness, we numerically compare the behavior of the stability of the FQH states in the LLL and 1LL as varying the strength of the in-plane B field and layer thickness. The stability of the FQH states are described by the ground state energy gap, wavefunction overlap and the static structure factor. We also introduce the effect of the Laudau level mixing, especially for the FQH states in 1LL.
The rest of this paper is arranged as following.  In Sec II, we review the single electron solution and analyze the PPs in different LLs with different parameters.  In Sec III, the numerical diagonalization for $\nu = 1/3$ and $\nu = 7/3$ states are implemented. The energy gaps as a function of the in-plane field at various parameters are compared.  The effect of the LL mixing is also introduced in the end of this section.  The summary and conclusion are given in Sec IV.

\section{Model and effective interaction}
 Without loss of generality, we assume that the in-plane B field is along $x$ direction. A general single particle Hamiltonian with a tilted magnetic field in a confinement potential can be written as
\begin{eqnarray}\label{h1}
H&=&\frac{1}{2m}\left[\left(P_x+eA_x\right)^2+\left(P_y+eA_y\right)^2+\left(P_z+eA_z\right)^2\right]\nonumber\\
&&+\frac{1}{2}m\omega_0^2z^2
\end{eqnarray}
where the second line gives the harmonic potential along $z$-axis which mimics the layer thickness of the two-dimensional electron gas (2DEG). The smaller $\omega_0$ means a larger thickness.  After defining the canonical momentum as $\pi_i=P_i+eA_i$, with $i=1,2,3$ along $x,y,z$ and $\pi_4=m\omega_0z$, the Eq.(\ref{h1}) can be written as
\begin{eqnarray}\label{h2}
H = \frac{1}{2m} \sum_{i=1}^{4} \pi_i^2
\end{eqnarray}
with the following commutation relations:
\begin{eqnarray}\label{commutation}
&&[\pi_1,\pi_2]=-i\ell_{B_z}^{-2},\nonumber\\
&&[\pi_2,\pi_3]=-i\ell_{B_x}^{-2},\nonumber\\
&&[\pi_3,\pi_1]=[\pi_1,\pi_4]=[\pi_2,\pi_4]=0,\nonumber\\
&&[\pi_3,\pi_4]=-i\ell_0^{-2},
\end{eqnarray}
where the three length scales are given by $\ell_{B_z}=1/\sqrt{eB_z},\ell_{B_x}=1/\sqrt{eB_x}$ and $\ell_0=1/\sqrt{m\omega_0}$. $\ell_0$ gives the characteristic width of the harmonic well.  The cyclotron energies in a magnetic field are defined by frequency $\omega_z = \frac{1}{m \ell_{B_z}^2}$ and $\omega_x = \frac{1}{m \ell_{B_x}^2}$. In order to diagonalize the Hamiltonian of Eq.~\ref{h2}, we used the Bogoliubov transformation~\cite{yang17prb} to write the Hamiltonian in the following form:
\begin{eqnarray}\label{h3}
H= \omega_1 X^\dagger X + \omega_2 Y^\dagger Y + \text{constant}.
\end{eqnarray}
where the new decoupled operators $(X, X^\dagger)$ and $(Y, Y^\dagger)$ are some linear combinations of the canonical momentum $\pi_is$.
  The single particle Hilbert space is thus built from these two sets of decoupled ladder operators, and the LLs are now indexed by two integers
\begin{eqnarray} \label{h4}
|m,n\rangle=\frac{1}{\sqrt{m!n!}}\left(X^\dagger\right)^m\left(Y^\dagger\right)^n|0\rangle,
\end{eqnarray}
where $|0\rangle$ is the vacuum state.  In the limit of $\omega_x\rightarrow 0$, when $\omega_0>\omega_z$, the operators $\left(X^\dagger, X\right)$ raises and lowers the in-plane LLs, while $\left(Y^\dagger,Y\right)$ raises and lowers the harmonic modes along z-axis (or the subbands).  The role of $X$ and $Y$ are reversed  for $\omega_0<\omega_z$. 

\begin{figure}
\includegraphics[width=1.0\linewidth]{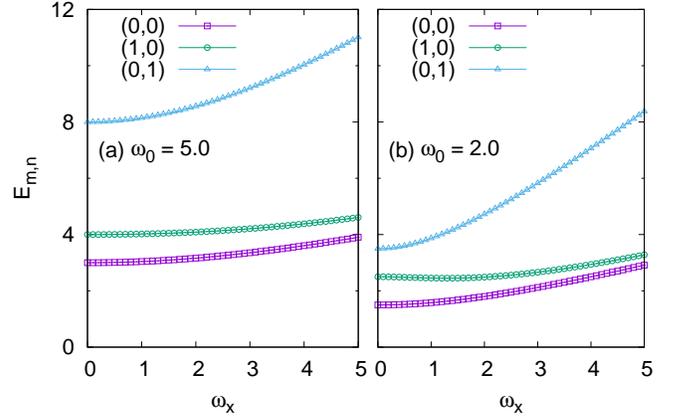}
\caption{\label{LLs}(Color online) The generalized LL energy as a function of the in-plane magnetic field $\omega_x$ for the lowest three LLs with different confinements (a) $\omega_0 = 5.0$ and (b)  $\omega_0 = 2.0$.  The index of the LL is labelled by $(m,n)$ in Eq.~\ref{h4}.}
\end{figure}

In Fig.~\ref{LLs}, we plot the energies of the lowest three generalized LLs as a function of $\omega_x$ for two different layer thicknesses with $\omega_0 = 5.0$ and $\omega_0 = 2.0$. For both cases, we find that the lowest two LLs are getting closing to each other while increasing $\omega_x$ and thus the in-plane B field mixes strongly the lowest two LLs. On the other hand, with comparing the results for two different $\omega_0$s, the larger layer thickness (smaller $\omega_0$) makes the third LL more closer to the other two LLs. It demonstrates that the effect of the LL mixing should be more profound for small $\omega_0$ and large $\omega_x$.

 With the exact solution of the single particle Hamiltonian, we can compute the form factor of the effective two-body interaction while projecting to the LLL. We now look at the full density-density interaction Hamiltonian with a bare Coulomb interaction
\begin{eqnarray}\label{hint}
H_{\text{int}}=\int d^3qV_{\vec q}\rho_q\rho_{-q},
\end{eqnarray}
where $V_{\vec q}=1/q^2$ is the Fourier components of the 3D Coulomb interaction, and $\rho_q=\sum_ie^{i\vec q\cdot r}$ is the density operator.  
The part relevant to the Landau level form factor is thus given by
\footnotesize
\begin{eqnarray}
&F_{m,n}&\left(\vec q,q_3\right)=\langle m,n|e^{i\left(q_1\tilde R^1+q_2\tilde R^2+q_3r^3\right)}|m,n\rangle\nonumber\\
&=&\langle m,n|e^{PX-P^*X^\dagger+QY-Q^*Y^\dagger}|m,n\rangle\nonumber\\
&=&e^{-\frac{1}{2}\left(PP^*+QQ^*\right)}\mathcal L_m\left(PP^*\right)\mathcal L_n\left(QQ^*\right)
\end{eqnarray}
\normalsize
where $\tilde R^1$, $\tilde R^2$ and $r^3$ are the coordinates of the cyclotron motion for electron in a B field.  By integrating out the component $q_3$, we then obtain the effective two-dimensional interaction.
\begin{eqnarray}\label{finalf}
V_{\vec q}^{\left(mn\right)}=\int_{-\infty}^{\infty}dq_3\frac{1}{|\vec{q}|^2+q_3^2}|F_{mn}(\vec q,q_3)|^2\label{effective2d}
\end{eqnarray}

 One should first note that Eq.(\ref{effective2d}) can be integrated exactly in the LLL (i.e. $m=n=0$). The result is as follows:
\footnotesize
\begin{eqnarray}\label{v00}
&&V_{\vec q}^{\left(00\right)}=\frac{1}{|q|}e^{-G_1\left(q_+,q_-\right)}G_2\left(q_+,q_-\right)\\
&&G_1\left(q_+,q_-\right)=\left(f_1^{12}+f_1^{21}\right)q_+q_-+\left(f_2^{12}+f_2^{21}\right)\left(q_+^2+q_-^{2}\right)\nonumber\\
&&G_2\left(q_+,q_-\right)=\pi\cos F_1e^{F_2}\nonumber\\
&&\qquad -i\sqrt\pi e^{\frac{F_1^2}{4F_2}}\left(\mathcal D\left(\frac{F_1-2iF_2}{2\sqrt{F_2}}\right)-\mathcal D\left(\frac{F_1+2iF_2}{2\sqrt{F_2}}\right)\right)\nonumber\\
&&F_1=|q|\left(q_++q_-\right)\left(f_3^{12}+f_3^{21}\right)\nonumber\\
&&F_2=|q|^2\left(f_4^{12}+f_4^{21}\right)\nonumber
\end{eqnarray}
\normalsize
in which $q_{\pm} = \frac{1}{\sqrt{2}}(q_1 \pm i q_2)$ and  $\mathcal D\left(x\right)$ is the Dawson integral. $f_i^{12}$ and $\widetilde{q}^{12}$ below are defined in Ref.~\onlinecite{yang17prb}. In the limit of infinitesimal sample thickness $\omega_0\rightarrow\infty$, $G_1\left(q_+,q_-\right)\rightarrow \frac{1}{2}\left(q_1^2+q_2^2\right), G_2\left(q_+,q_-\right)\rightarrow 1$. 
For the higher LLs, we calculate the effective interaction for LL (1,0) and (0,1):
\footnotesize
\begin{eqnarray}
V_{\vec{q}}^{10} &=& \int_{-\infty}^{\infty} d q_{3} \frac{e^{-(\widetilde{q}^{12}+\widetilde{q}^{21})}}{|\vec{q}|^2+q_3^2}[1+(\widetilde{q}^{12})^2-2\widetilde{q}^{12}]		\\
V_{\vec{q}}^{01} &=& \int_{-\infty}^{\infty} d q_{3} \frac{e^{-(\widetilde{q}^{12}+\widetilde{q}^{21})}}{|\vec{q}|^2+q_3^2}[1+(\widetilde{q}^{21})^2-2\widetilde{q}^{21}].
\end{eqnarray}
\normalsize
By using the following identities: 
\footnotesize
\begin{eqnarray}
I_0(a,b) &=& \int_{-\infty}^{\infty} d x \frac{e^{-ax-bx^2}}{1+x^2}=\frac{1}{2}\pi e^{b-ia} \times \nonumber	\\ &&
[e^{2ia} \text{Erfc}(\frac{2b+ia}{2\sqrt{b}})+ \text{Erfc}(\frac{2b-ia}{2\sqrt{b}})] 	\nonumber\\
I_1(a,b) &=& \int_{-\infty}^{\infty} d x \frac{x e^{-ax-bx^2}}{1+x^2}=\frac{1}{2}i\pi e^{b-ia} \times \nonumber	\\ &&
[-e^{2ia} \text{Erfc}(\frac{2b+ia}{2\sqrt{b}})+ \text{Erfc}(\frac{2b-ia}{2\sqrt{b}})] 	\nonumber\\
I_2(a,b) &=& \int_{-\infty}^{\infty} d x \frac{x^2 e^{-ax-bx^2}}{1+x^2}=\frac{\sqrt{\pi}}{\sqrt{b}} e^{\frac{a^2}{4b}} -I_0(a,b) 		\nonumber\\
I_3(a,b) &=& \int_{-\infty}^{\infty} d x \frac{x^3 e^{-ax-bx^2}}{1+x^2}=-\frac{\sqrt{\pi}ae^{\frac{a^2}{4b}}}{2b^{3/2}}  -I_1(a,b) \nonumber \\
I_4(a,b) &=& \int_{-\infty}^{\infty} d x \frac{x^4 e^{-ax-bx^2}}{1+x^2}\nonumber \\
&=&-\frac{\sqrt{\pi}(a^2+2b(1-2b))e^{\frac{a^2}{4b}}}{4b^{5/2}} +I_0(a,b) \nonumber
\end{eqnarray}
\normalsize
The $V_{\vec{q}}^{10}, V_{\vec{q}}^{01}$ are obtained as follows:
\footnotesize
\begin{eqnarray}
V_{\vec{q}}^{10} &=&  [1+(0.5f_1^{12}|q|^2)^2+(f_2^{12}(q_1^2-q_2^2))^2+2f_1^{12}|q|^2f_2^{12}  \nonumber \\
&\times&(q_1^2-q_2^2)^2  - f_1^{12}|q|^2 -2f_2^{12}(q_1^2-q_2^2)^2] V_{\vec{q}}^{00} + 2\sqrt{2}q_1f_3^{12}  \nonumber \\	
&\times& (0.5f_1^{12}|q|^2+f_2^{12}(q_1^2-q_2^2)^2-1.0) e^{-G_1(q_{+},q_{-})}  I_1(F_1,F_2)   \nonumber \\	
&+&	[(\sqrt{2}q_1f_3^{12})^2 + f_1^{12}|q|^2 f_{12}^4 + 2f_2^{12}(q_1^2-q_2^2)f_{12}^4  -(f_{12}^4)^2)]|q|   \nonumber \\
&\times& e^{-G_1(q_{+},q_{-})}	 I_2(F_1,F_2) +2\sqrt{2}q_1f_3^{12}f_{12}^4 |q|^2 e^{-G_1(q_{+},q_{-})}  \nonumber \\
&\times&I_3(F_1,F_2)+ (f_{12}^4)^2 |q|^3 e^{-G_1(q_{+},q_{-})} I_4(F_1,F_2)
\end{eqnarray}
\normalsize
$V_{\vec{q}}^{01}$ is obtained by switching the $\widetilde{q}^{12}$ by $\widetilde{q}^{21}$.

 For a two-body interaction without rotational symmetry, some of us recently found~\cite{yang2} that a generalized pseudopotential description can be defined by:
\begin{eqnarray} \label{PPS}
 V^+_{m,n}(\bold{k}) &=& \lambda_n \mathcal{N}_{mn}(L_m^n(|k|^2) e^{-|k|^2/2} {\bold{k}}^n + c.c) \nonumber \\
 V^-_{m,n}(\bold{k}) &=& -i \mathcal{N}_{mn}(L_m^n(|k|^2) e^{-|k|^2/2} {\bold{k}}^n - c.c)
\end{eqnarray}
where the normalization factors are $\mathcal{N}_{mn} = \sqrt{2^{n-1}m!/(\pi(m+n)!)}$ and $\lambda_n = 1/\sqrt{2}$ for 
$n = 0$ or $\lambda_n = 1$ for $n \neq 0$. They satisfy the orthogonality
\begin{eqnarray} 
 \int V^\sigma_{m,n}(\vec{k}) V^{\sigma'}_{m',n'}(\vec{k}) d^2 k  = \delta_{m,m'}\delta_{n,n'}\delta_{\sigma,\sigma'}
\end{eqnarray}
thus the effective two-body interaction including the anisotropic ones can be expanded as
\begin{eqnarray} \label{ppformula}
 V_{\text{eff}}(\bold{k}) = \sum_{m,n,\sigma}^\infty c^{\sigma}_{m,n} V^{\sigma}_{m,n}(\bold{k})
\end{eqnarray}
with the coefficient
\begin{eqnarray}
 c^{\sigma}_{m,n} = \int d^2 k V_{\text{eff}}(\bold{k})  V^{\sigma}_{m,n}(\bold{k}).
\end{eqnarray}

\begin{figure}
\includegraphics[width=8cm]{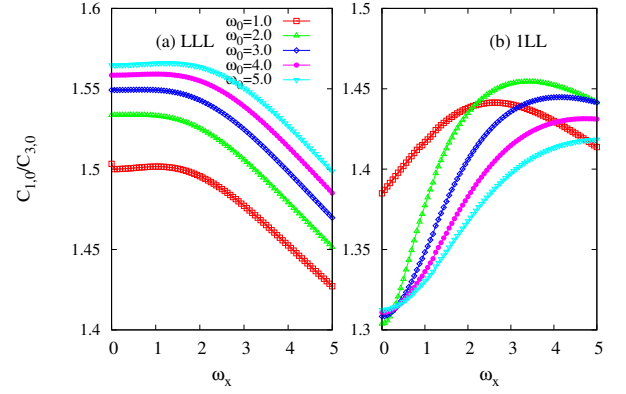}
\caption{\label{pps}The PPs $c_{1,0}/c_{3,0}$ as a function of $\omega_x$ for different $\omega_0$s on the LLL (a) and 1LL (b).  In the LLL, the ratio $c_{1,0}/c_{3,0}$ monotonically decreases with increasing the in-plane B field. On the contrary, the ratio increases and reaches a maximum at some  specific $\omega_x$.}
\end{figure}
  
In Fig.~\ref{pps}, we plot the ratio of the first two dominant pseudopotentials $c_{1,0}/c_{3,0}$ as a function of the in-plane B field $\omega_x$ with different thicknesses $\omega_0$. It is interesting to see that the $c_{1,0}/c_{3,0}$ in the LLL monotonically decreases as increasing the $\omega_x$; however, things are different in the 1LL. The ratio increases for small tilting and reaches its maximum at some specific value of $\omega_x$ before decreasing at large tilting. From the analysis of the PPs, we therefore have a conjecture that the Laughlin state at $\nu = 1/3$ is smoothly destablized by the in-plane B field. However, in the 1LL, the FQH state at $\nu = 7/3$ or $\nu = 8/3$ will be stabilized with a small tilted field and finally destroyed by a large tilting field.

\section{Numerical diagonalization}
\begin{figure}
\includegraphics[width=1.0\linewidth]{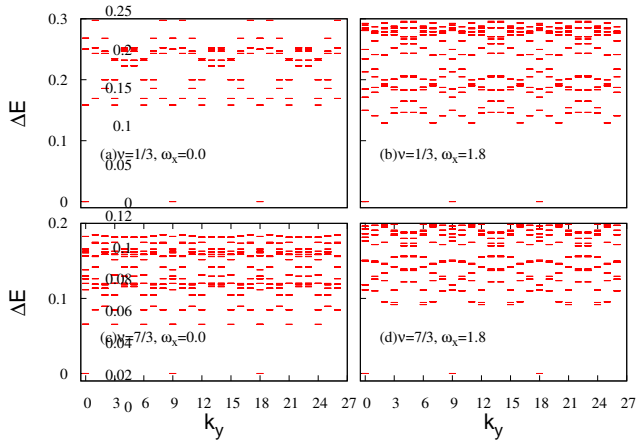}
\caption{\label{spectrum} Energy spectrum for 9 electrons at $\nu = 1/3$ and $\nu = 7/3$ without and with a small tilted field. The three-fold degeneracy of the ground states reveals the Laughlin type of the ground state. With a small tilted field at $\omega_x = 1.8$, the gap of the $1/3$ and $7/3$ behaviors differently with comparing to the untilted case. The confinement strength is set to be $\omega_0 = 2.0$.}
\end{figure}

In this section, we systematically study the FQH state at $\nu = 1/3$ on the LLL and 1LL in a torus geometry with the effect of the tilted magnetic field. Since the particle hole symmetry of the two-body Hamiltonian, the energy spectrum at $\nu =2/3(8/3)$ is the same as that for $\nu = 1/3(7/3)$. We therefore only consider the FQH state at $\nu = 1/3$ in LLL and $\nu = 7/3$ in 1LL.  In the following, we set $\omega_z = 1$ for simplicity. In Fig.\ref{spectrum}, we plot the energy spectrum for 9 electrons at $1/3$ filling on different LLs without and with the tilted magnetic field. The three-fold degeneracy and large $c_{1,0}$ PPs as shown in Fig.~\ref{pps} on both LLs demonstrate that the Laughlin state describes the ground state very well. A small tilting of the magnetic field does not break the ground state degeneracy except varying the energy gap.  However, when we comparing the energy spectrums between the cases with $\omega_x = 0$ and $\omega_x = 1.8$, it is interesting to see that the gap of the $1/3$ state is reduced but the gap of the $7/3$ state is enhanced by this in-plane field.

\begin{figure}
\includegraphics[width=1.0\linewidth]{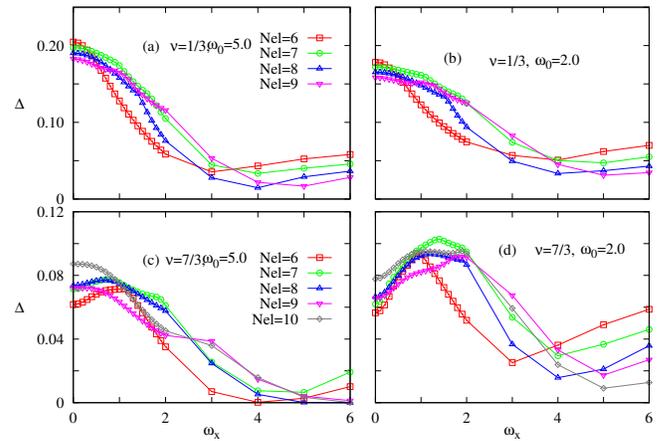}
\caption{\label{gap} The energy gap as a function of the in-plane magnetic field. For $1/3$ state, the gap always decreases as increasing $\omega_x$ for $\omega=5.0$ (a) and $\omega = 2.0$ (b).  However, the gap for the $7/3$ state has different behaviors at different $\omega_0$s. For $\omega_0 = 5.0$ (c), we find the energy gap for 9 and 10 electrons still monotonically decreases as increasing $\omega_x$(We assume that the increments for small systems is due to the finite size effect). When $\omega_0 = 2.0$ (d), the gap increases for small tilting and decreases for large tilting.}
\end{figure}

As being indicated by the PPs  and the energy spectrum, we compare the energy gap of the $1/3$ and $7/3$ FQH states as a function of the in-plane B field $\omega_x$. The energy gap is defined as the energy difference between the ground state and the lowest excited state (which corresponds to the minimal energy of the magneto-roton excitation) in the spectrum. The results are shown in Fig.~\ref{gap}. We compare the behavior of the energy gap as a function of the in-plane B field for two different $\omega_0$s.  The energy gap of the $1/3$ state always monotonically decays as increasing $\omega_x$ as shown in Fig.~\ref{gap}(a) and (b). However,  the gap for $7/3$ state has different tendencies for different $\omega_0$s. When $\omega_0 = 5.0$ as shown in Fig.~\ref{gap}(c), similar to the LLL, the gap for the largest two systems (9 and 10 electrons) still monotonically decreases as increasing $\omega_x$; however, for the case of $\omega_0 = 2.0$, we find the energy gap increases and reaches to its maximum for  a small $\omega_x$ and deminishes for a large $\omega_x$.

\begin{figure}
\includegraphics[width=8cm]{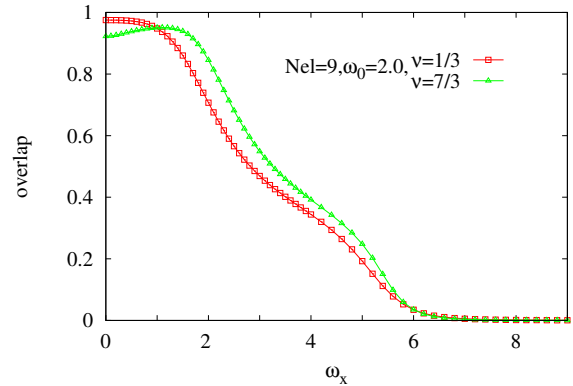}
\caption{\label{overlap} The wavefunction overlap between the ground state and the Laughlin state as a function of $\omega_x$ at $\omega_0 = 2.0$. Similar to the energy gap, we find the overlap for the $1/3$ state monotonically decreases as increasing the in-plane B field. However, the overlap for $7/3$ state increases for small $\omega_x$ and decays to zero for large tilting.}
\end{figure}

Besides the energy gap, we also consider the wavefunction overlap under the effects of the in-plane magnetic field. Here we use the exact Laughlin state as the reference wavefunction which is obtained by diagonalizing the $V_1$ Hamiltonian. Fig.~\ref{overlap} shows that the overlap between the $1/3$ ground state and the Lauglin state monotonically decays as increasing $\omega_x$ when $\omega_0 = 2.0$. However, for the ground state of $\nu = 7/3$, the overlap is enhanced for small tilting which is consistent to the energy gap calculation.

\begin{figure}
\includegraphics[width=8cm]{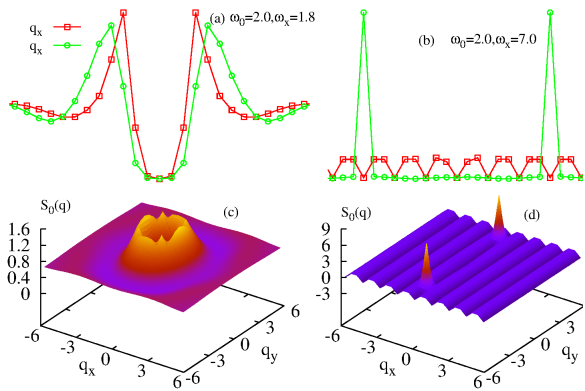}
\caption{\label{sf} Static structure factor for 9 electrons at $\nu$=7/3 with $\omega_0 = 2.0$. (a) (b) are the lateral view of (c) and (d).
With a small in-plane field, the system has a strong anisotropy (c). After a phase transition for large $\omega_x$, the system enters into a charge-density-wave-like compressible phase which is characterized by two peaks in the $S_0(q)$ as shown in (d). }
\end{figure}

The symmetry breaking and phase transition of the FQH states can be described by the projected static structure factor, which is defined as\cite{edtilt}
\begin{eqnarray}
	S_0(q) = \frac{1}{N_{el}}\langle 0 \vert \sum_{i, j}e^{iq\cdot (R_i-R_j))}\vert 0 \rangle
\end{eqnarray}
Where $\vert 0 \rangle $ is the ground state and $R_i$ is the guiding center coordinate of the $i$'th particle.  In Fig. \ref{sf}, we plot the static structure factor of the $7/3$ state with different in-plane magnetic fields.  When the in-plane field is small, i.e., $\omega_x = 1.8$ as shown in Fig. \ref{sf}(c) and its lateral view (a), the system has strong anisotropy between $q_x$ and $q_y$ directions.  With a strong in-plane field $\omega_x = 7.0$, as explained in Fig.~\ref{gap}, a phase transition has been occurred and the system enters into a compressible state. From the Fig. \ref{sf} (b) and (d), we find there are only two peaks in the structure factor which characterizes the ground state is a charge density wave (CDW) state in a large tilted field.
 
As shown in Fig.~\ref{LLs}, the lowest two LLs are very close to each other and the in-plane magnetic field reduces the gap of the two lowest LLs. Therefore, the neglection of the LL mixing in previous calculation may not be a good approximation. Generally, the strength of the LL mixing is defined as the ratio of characteristic Coulomb interaction and kinetic energy in the magnetic field:
\begin{eqnarray}
  \kappa \equiv \frac{e^2}{\hbar \omega_z \epsilon l_B}
\end{eqnarray}
The Landau level mixing is a virtual excitation process of the electrons hopping between the occupied and unoccupied LLs.  Recently, several research groups~\cite{nayak, Nayak_2013, Ed13, MacDonald13} calculated the LL mixing corrections on the PPs in a perturbation way. The mainly contribution of the LL mixing can be classified into a correction on the two-body and three-body interactions. Here we consider the effects of the LL mixing in the tilted magnetic field system.  With a magnetic field in $x-z$ plane, we can generally decompose the two-body Hamiltonian into $x$ and $z$ components:
\begin{eqnarray}
	\mathcal{H} = \mathcal{H}_z + \mathcal{H}_x.
\end{eqnarray}
In the language of the PPs, the first term contributes the diagonal PPs $V_{m,m}$ and the second term contributes the off-diagonal PPs $V_{m,n\neq m}$.  In the following, we consider the LL mixing correction on the diagonal part of the two-body interaction:
\begin{eqnarray}
H_z = \mathcal{H}_z + \lambda \frac{1}{1+\omega_x^2} H_{LLM}
\end{eqnarray}

\begin{figure}
\includegraphics[width=1.0\linewidth, height = 5cm]{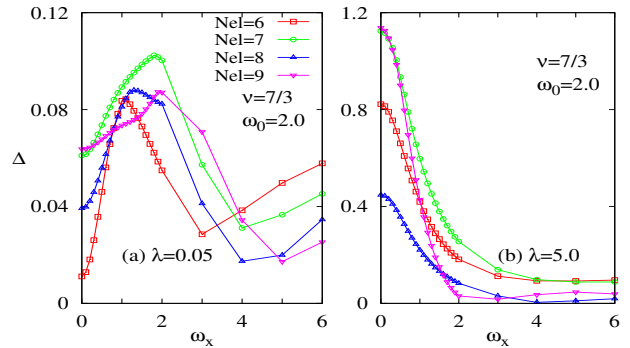}
\caption{\label{llm} The energy gap after considering the effect of the LL mixing on the two-body interaction. When LL mixing is small, the gap still has a bump as increasing $\omega_x$. In the case of strong LL mixing, the bump disappears and the gap monotonically decreases as that in the LLL.}
\end{figure}

In our model, the strength of the confinement $\omega_0$ in $z$ direction can't be converted to the layer thickness directly. Therefore we approximately use the two-body LL mixing correction with a typical layer thickness $l_B$\cite{nayak, Nayak_2013, Ed13, MacDonald13} in $H_{LLM}$. The parameter $\lambda$ describes the strength of the LL mixing which has the same role of the $\kappa$.   Here we have a factor $\frac{1}{1+\omega_x^2}$  which is the proportion of the kinetic energy in $z$ direction (because of $\omega_z = 1$). We should note that the correction is calculated in isotropic system, so the way of introducing the LL mixing is just an approximation.  The results are depicted in Fig.~\ref{llm}. We set the system parameters at $\omega_0 = 2.0$ for $\nu = 7/3$ state. As shown in Fig.~\ref{gap} (d), this parameter corresponding to the case of increasing the gap by small tilting.  When the LL mixing is small, as shown in Fig.~\ref{llm} (a) with $\lambda = 0.05$, we find the ground state energy gap still has a bump as a function of the $\omega_x$.  However, when the LL mixing is strong enough, such as $\lambda = 5.0$ as shown in Fig.~\ref{llm}(b), the bump disappears and the gap monotonically reduced by the in-plane field as that in the LLL.   Therefore, we conclude that in a strong LL mixing, the stability of the $7/3$ state is no longer enhanced by the in-plane field.
 
\section{Summaries and discussions}
 In conclusion, we systematically study the stability of the FQH state at $\nu = 1/3$ and $\nu = 7/3$ with the effects of the in-plane magnetic field. By exactly solving the single particle Hamiltonian in a tilted magnetic field and harmonic confinement along $z$ direction, we obtain the effective two-body interaction in the lowest three Landau levels.  With expanding these effective interaction by the generalized pseudopotentials, we find the $c_{1,0}/c_{3,0}$ behaviors differently in the LLL and 1LL which indicating different behaviors under tilting.  The results of the numerical exact diagonalization are consistent with the PPs analysis.  By comparing the ground state energy gap and the wavefunction overlap, we conclude that the stability of the FQH at $1/3$ is monotonically reduced by the in-plane B field. However, for the $7/3$ FQH state on the 1LL, we find a small tilting magnetic field can stabilize the state when $\omega_0$ is small, such as increasing the energy gap and wavefunction overlap.  
From the calculation of the static structure factor, we observe the anisotropy of the system for small tilting and finally a phase transition into a CDW-like state occurs in large tilting. Our numerical results are qualitatively consistent to that of the experimental observations~\cite{dean08,chi12} in which the enhancements of the FQH on the 1LL were indeed observed in a small in-plane magnetic field. However, with considering the effect of the LL mixing correction on the two-body interaction, we find that a strong LL mixing correction can diminish and finally erase the enhancements of the gap. Therefore, we conjecture that the LL mixing should be small in those experiments.

Here we should note that we just consider the two-body correction of the LL mixing in our calculation. The two-body interaction does not break the particle-hole symmetry of the system. Therefore, all the results for $1/3$($7/3$) are the same as that for $2/3$(8/3) state. The breaking down of the particle-hole symmetry needs the three-body terms of the LL mixing which will be included in future study.

\begin{acknowledgements}
This work was supported by National Natural Science Foundation of China Grants No. 11674041, No. 91630205 and Chongqing Research Program of Basic Research and Frontier Technology Grant No. cstc2017jcyjAX0084.  
\end{acknowledgements}


\begin{thebibliography}{99}

\bibitem{tsui}  D. C. Tsui, H. L. Stormer, and A. C. Gossard, Phys. Rev. Lett. {\bf 48}, 1559 (1982).
\bibitem{prange} R. Prange and S. Girvin, The Quantum Hall effect, Graduate texts in contemporary physics (Springer- Verlag, 1987), ISBN 9783540962861
\bibitem{haldane1} F. D. M. Haldane, Phys. Rev. Lett. {\bf 51}, 605 (1983).
\bibitem{haldane2} F. D. M. Haldane, The Quantum Hall Effect (Springer, New York, 1990).
\bibitem{laughlin} R. B. Laughlin, Phys. Rev. Lett. {\bf 50}, 1395 (1983).
\bibitem{MR} G. Moore and N. Read, Nucl. Phys. B {\bf 360}, 362 (1991).
\bibitem{Wilczek91} M. Greiter, X.-G. Wen, and F. Wilczek, Phys. Rev. Lett. {\bf 66}, 3205 (1991).
\bibitem{lilly99} M. P. Lilly, K. B. Cooper, J. P. Eisenstein, L. N. Pfeiffer, and K. W. West, Phys. Rev. Lett. {\bf 82}, 394 (1999).
\bibitem{du99} R. R. Du, D. C. Tsui, H. L. Stormer, L. N. Pfeiffer, K. W. Baldwin, and K. W. West, Solid State Commun. {\bf 109}, 389 (1999).
\bibitem{dean08} C. R. Dean, B. A. Piot, P. Hayden, S. Das Sarma, G. Gervais, L. N. Pfeiffer, and K. W. West, Phys. Rev. Lett. {\bf 101}, 186806 (2008).
\bibitem{chi12} G. T.  Liu, C. Zhang, D. C. Tsui, I. Knez, A. Levine, R. R. Du, L. N. Pfeiffer, and K. W. West, Phys. Rev. Lett. {\bf 108},196805 (2012).
\bibitem{Cooper08} N. R. Cooper, Adv. Phys. {\bf 57}, 539 (2008).
\bibitem{qiu11} R. Z. Qiu, S. P. Kou, Z. X. Hu, X. Wan and S. Yi, Phys. Rev. A {\bf 83}, 063633 (2011).
\bibitem{haldane3} F. D. M. Haldane, Phys. Rev. Lett. {\bf 107}, 116801 (2011).
\bibitem{yang2} B. Yang, Z. X. Hu, C. H. Lee and Z. Papic, Phys. Rev. Lett. {\bf 118}, 146403 (2017).
\bibitem{hu18} Z-X. Hu, Q. Li, L-P. Yang, W-Q. Yang, N. Jiang, R-Z. Qiu and B. Yang, Phys. Rev. B {\bf 97}, 035140 (2018).
\bibitem{zpapic13} Z. Papic, Phys. Rev. B {\bf 87}, 245315 (2013).
\bibitem{yang1} B. Yang, Z. Papic, E.H. Rezayi, R. N. Bhatt and F.D.M. Haldane, Phys. Rev. B {\bf 85}, 165318 (2012).
\bibitem{yang17prb} B. Yang, C-H. Lee, C. Zhang, and Z-X. Hu, Phys. Rev. B, {\bf 96} 195140 (2017).
\bibitem{edtilt} E. H. Rezayi and F. D. M. Haldane, Phys. Rev. Lett. {\bf 84}, 4685 (2000).
\bibitem{nayak} W. Bishara and C. Nayak, Phys. Rev. B {\bf 80}, 121302 (2009).
\bibitem{Nayak_2013} M. R. Peterson and C. Nayak, Phys. Rev. B {\bf 87},245129 (2013).
\bibitem{Ed13}S. H. Simon and E. H. Rezayi, Phys. Rev. B {\bf 87}, 155426, (2013).
\bibitem{MacDonald13} I. Sodemann and A. H. MacDonald, Phys. Rev. B {\bf 87}, 245425 (2013)

 





 

 
 

 \end{thebibliography}
\end{document}